\documentclass[a4paper,11pt]{article}
\usepackage{jinstpub} 
\usepackage{lineno}
\usepackage{multirow}

\title{Functional Verification for Endcap Concentrator ASICs in the High-Granularity Calorimeter Upgrade of CMS}

\author[a,1]{M.~Lupi,\note{Corresponding author.}}
\author[a]{G.~Bergamin,}
\author[a]{D.~Ceresa,}
\author[b]{D.~Coko,}
\author[c]{G.~Cummings,}
\author[c]{V.~Gingu,}
\author[d]{M.~Hammer,}
\author[c]{J.~Hirschauer,}
\author[c]{J.~Hoff,}
\author[c]{N.~Kharwadkar,}
\author[a]{S.~Kulis,}
\author[c]{C.~Mantilla-Suarez,}
\author[c]{D.~Noonan,}
\author[c]{P.~Rubinov,}
\author[a]{S.~Scarf\'i,}
\author[c]{A.~Shenai,}
\author[c]{C.~Syal,}
\author[c]{X.~Wang,}
\author[c]{R.~Wickwire,}
\author[e]{and J.~Wilson}

\affiliation[a]{CERN, 1, Esplanade des Particules, Meyrin, Switzerland}
\affiliation[b]{University of Split, R.  Bo\v skovi\' ca 32 Split, Croatia}
\affiliation[c]{FNAL, Batavia IL 60510, USA}
\affiliation[d]{Argonne National Laboratory, Lemont IL 60439, USA}
\affiliation[e]{Baylor University, Waco TX 76706, USA}

\collaboration[c]{on behalf of CMS collaboration}

\emailAdd{matteo.lupi@cern.ch}

\abstract{
The High-Granularity Calorimeter (HGCAL) will replace the current CMS Endcap Calorimeter during Long-Shutdown 3. 
The Endcap Concentrator (ECON) ASICs represent key elements in the readout chain, processing trigger (ECON-T) and data (ECON-D) streams from the HGCROC to the lpGBT. 
The ECONs will operate in a radiation environment with a High-Energy Hadron (${E\geq20MeV}$) flux up to $2\cdot10^{7} cm^{-2}s^{-1}$.

This contribution describes the Universal Verification Methodology (UVM)-based functional verification of the ECON ASICs focusing on the re-use of existing components to manage the complexity of the verification environment.
}


\arxivnumber{2411.03476} 

\begin{document}
\maketitle
\flushbottom

\section{Introduction}\label{sec:intro}
The High Granularity Calorimeter (HGCAL) is a 47-layer sampling calorimeter that will operate in the CMS detector, upgraded for the High-Luminosity (HL) LHC~\cite{tdr:hgcal}.
The detector is divided into two parts, one using silicon sensors and one using scintillating tiles as active material.
The silicon section will consist of section will consist of over $600 m^2$ of hexagonal-shaped sensors, with over six million channels.
The scintillating tile section will cover $370 m^2$, with around 280 thousand channels.
Both sections will be read out using two flavours of the HGCROC Front-End (FE) ASIC~\cite{hgcroc2022}.
Subsequently, the digitized data undergo processing and zero-suppression along two independent paths, managed by two Endcap Concentrators (ECONs): the trigger path, by the ECON-T ASIC, and the data path, by the ECON-D ASIC~\cite{econ2024}.
The former produces Level-1 Trigger (L1) primitives for each bunch crossing, whereas the latter forwards data packets to the acquisition system at an average L1 rate of 750 kHz. 
These ASICs operate in tandem, sharing a foundational infrastructure, synchronized clocking mechanisms, and input/output protocols, while drawing upon established silicon-proven Intellectual Property (IP). 
Both ECONs will operate in a radiation environment with a High-Energy Hadrons (${E\geq20MeV}$) flux up to $2\cdot10^{7} cm^{-2}s^{-1}$. 
Distributed Triple Modular Redundancy (TMR) and Error Correction Codes (ECCs) are implemented to allow the ECONs to operate in the presence of Single-Event Effects (SEEs).

\begin{figure}[thbp]
\centering
\includegraphics[width=.8\textwidth]{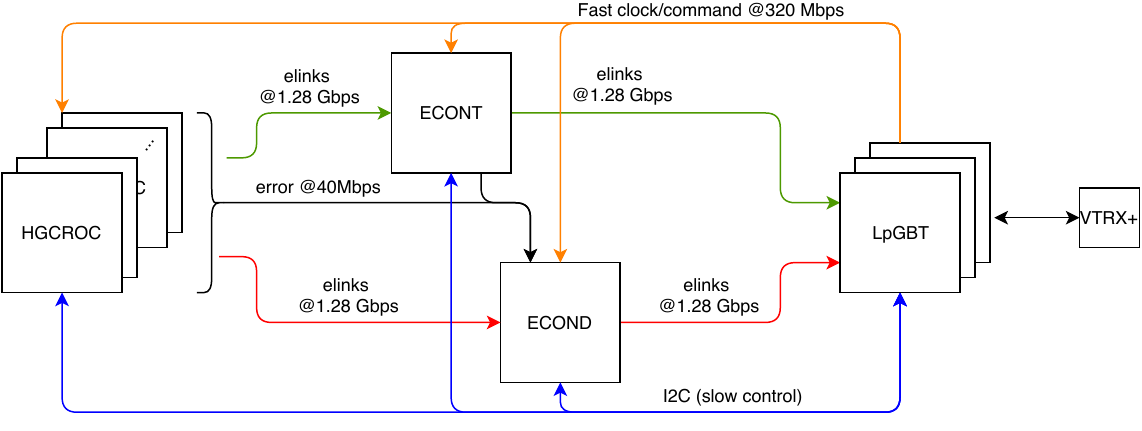}
\caption{Example of generic detector module with multiple lpGBTs, HGCROCs and two ECONs.\label{fig:module}}
\end{figure}

The whole detector will employ more than one hundred thousand HGCROCs and more than twenty thousand of each ECON-D and ECON-T.
The detector modules, an example shown in figure~\ref{fig:module}, will differ, among other parameters, for the number of HGCROCs associated with each ECON and for the number of lpGBTs~\cite{lpgbt2023} used for the data and trigger uplinks.
High-density modules, with $0.5 cm^2$ cells, will have up to six HGCROCs per ECON, whereas Low-density modules, with $1.2 cm^2$ cells, will have up to three HGCROCs per ECON.
Each HGCROC has two high-speed data (to ECON-D) links and four high-speed trigger (to ECON-T) links.
The number of high-speed links used in each ECON is configurable in the range of one to six, for ECON-D, and one to thirteen for ECON-T.
In most of the detector areas, the HGCROCs and ECONs will share the I2C bus, used to configure the ASICs, and will receive the clock and fast control signals at the same time.

\section{Verification strategy}
During the initial phases of the ECONs design, the verification team was tasked with developing a verification plan. 
The verification plan is composed of two main parts, a set of verification goals, i.e. features or scenarios that need to be verified, and a verification strategy, i.e. how the verification plan will be implemented to achieve the aforementioned goals.
The first translates into a set of test scenarios and coverage coding, i.e. a set of \emph{Systemverilog} \emph{covergroups} and \emph{coverpoints}~\cite{ieee:sv:lrm}, which is ASIC specific and is difficult to re-use across different projects.
The latter can easily be replicated in different designs.
The verification strategy for ECON-D and ECON-T was designed to target the constraints coming from topologies of the heterogeneous detector modules and from the radiation environment.
At the same time, the verification strategy should account for similarities between the two ECONs during the design of the verification components.

Both ECON-D and ECON-T were verified using multiple, complementary approaches: a strict Continuous Integration (CI) embedded into the version control system, a functional verification framework based on Universal Verification Methodology (UVM)~\cite{ieee:uvm}, and a set of formal verification techniques.

The ECONs were developed using \emph{git} as a version control system and \emph{GitLab} as a development platform.
A CI was setup with a gate-keeping function: at each change request, the code is linted and checked, making sure all the verification and implementation collaterals can be regenerated. 
For example to ensure that the RTL can correctly be triplicated with TMRG~\cite{tmrg2017} and that it can be elaborated by the simulator.
Special care was taken to ensure that the simulation results were repeatable and that failing test cases were reproducible.
The usage of \emph{git} tags allows uniquely identifying the RTL and verification code which is used to generate a failing test case.

UVM is, de facto, the industry standard for functional verification.
It implements coverage-driven constrained-random verification.
Moreover, UVM encourages the re-use of verification components (UVCs).
Given the similarities between the input and output stages of both ECONs, re-usability was exploited in the development of verification components.
The UVM testbenches were used as the main component to verify the compliance of the two ECONs to the detector requirements and the designer's specifications.

Formal verification, in particular model checking, is based on techniques which can prove certain conditions or provide a counter-example, if it exists.
In comparison, functional verification proves a design through repeated simulation of test cases whereas formal verification proves a design mathematically.
In the verification of the two ECONs, formal verifications techniques were mainly used to prove the correct implementation of TMR using techniques and they are described in~\cite{vrf:frm2023}.
The aforementioned techniques cannot demonstrate SEU tolerance by itself.
However, informal analysis of the functional verification of SEU tolerance across multiple HEP ASIC designs revealed that the majority of time spent by verification teams is actually spent in uncovering bugs that ultimately originate in how the RTL is written (using a non-voted signal in place of a voted one, a missing TMRG pragma~\cite{tmrg2017}, etc.).
These techniques allow providing the verification teams RTL in which properly structured TMR has been assured, thereby greatly improving the efficiency of the functional verification of SEU tolerance.

\section{Testbench architecture and vertical re-use}
The testbench architecture was designed to achieve the verification goals.
Three types of testbenches were developed: block-, ASIC-, and system-level testbenches.

Block-level testbench was used whenever the block was considered difficult to cover in the ASIC-level testbench. 
Specifically, in the verification of the ECON-D formatter, the Design Under Test (DUT) is a block with a relatively big fan-in, that receives data and formats them in real-time.
The block needs to process both in- and out-of-order packets with interrupts, i.e. a request to drop the current packet, arriving at any time.
Moreover, the block receives input data, stored in non-triplicated SRAMs, where only part of the data is protected with ECC.
During the ECON-D operation, SEUs will corrupt data stored in the SRAMs and the DUT will have to partially correct them.
A dedicated testbench for this DUT allowed emulating input data affected by SEUs and verifying that the DUT correctly treats these data.

ASIC-level testbenches, one for each ECON, allow reaching most of the verification goals.
The clocks and fast commands are provided through a dedicated UVC which accounts for jitter, the launching edge of the signals, and skew between clocks and data, among other things.
The ASIC is stimulated using two different data sources in input: data from physics simulations, stored in Comma-Separated Value (CSV) files, or randomized data generated by a HGCROC emulator.
The first allows running simulated data through the ECONs; the latter allows testing for corner cases and a wide range of scenarios covering the potential use cases of the ASICs in different locations in the detector.
The HGCROC emulator also allows the generation of HGCROC data with SEUs, mimicking a real operational scenario.  This allows for the testbench to see if the ECONs can cope with inconsistent or partially corrupted data while being able to flag these to the back-end.
One of the tasks of both ECONs is to be able to temporally align the incoming HGCROC data for parallel processing.
A dedicated UVC was written to randomize the delay with which each ECON channel receives the HGCROC data.
This allows the testbench to account for different PCB traces from the HGCROCs to the ECONs in different locations of the detector.

\begin{figure}[thbp]
\centering
\includegraphics[width=.99\textwidth]{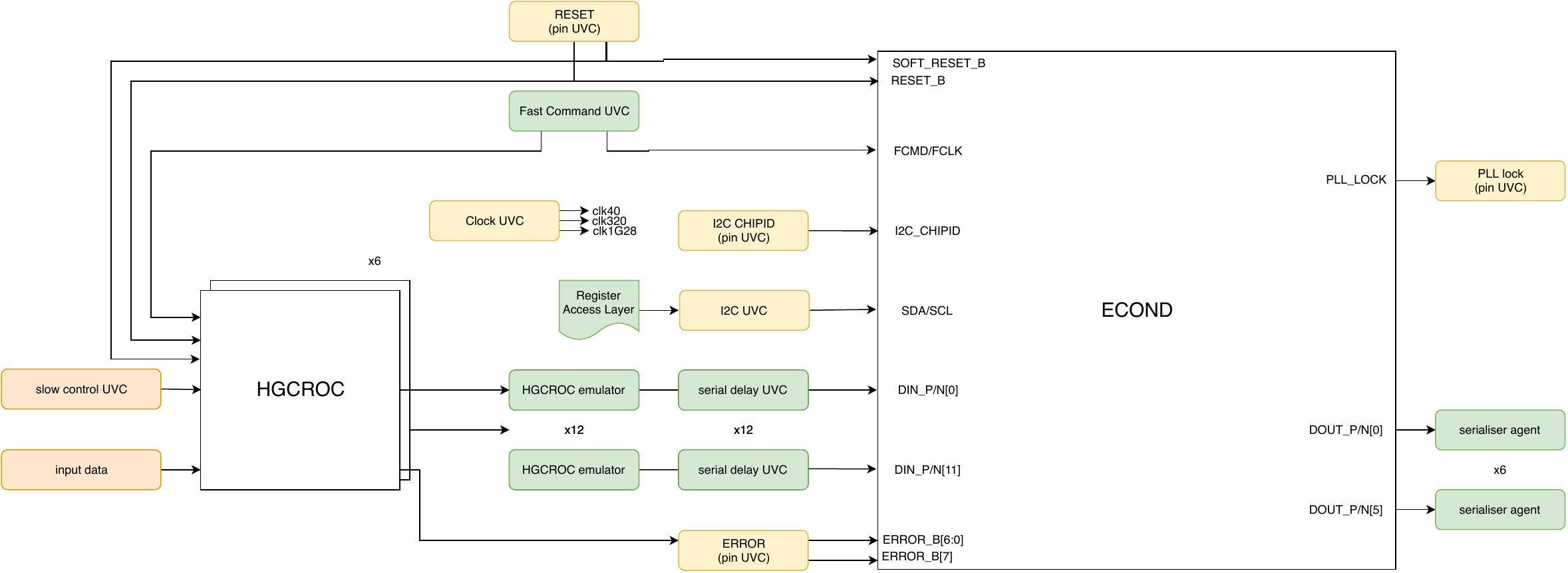}
\caption{Example of system-level testbench for ECOND. 
The testbench re-uses multiple re-usable UVCs, in yellow, and some UVCs developed specifically for the ECONs verification. 
These are re-usable in both ECONs testbenches. 
In orange there are the UVCs developed specifically for the HGCROC.\label{fig:tbsys}}
\end{figure}

System level testbenches, one for each ECON, allow verification that the ASICs are operating correctly and in sync with the HGCROCs.
An example of the ECON-D testbench is shown in figure~\ref{fig:tbsys}.
In particular, both ECONs and the HGCROC receive clocks and fast commands at the same time and they need to execute certain operations in sync, e.g. temporal alignment or data pipeline flushing.
Even if these testbenches require a non-negligible investment in development and execution time, they allow co-simulating the RTL of the HGCROC and the ECONs verifying that the specifications of both ASICs are compatible.

When comparing the three different levels of testbenches it is important to keep in mind a few key factors.
A block-level testbench is much faster to bring up, run tests on, and debug compared to a system-level testbench.
At the same time, it is important to focus on ASIC- or system-level scenarios as they are more likely to emulate realistic scenarios under which the detector will operate.

Vertical re-use of UVCs played a key role in developing the testbenches: each of the verification components was designed with re-use in mind and considering how it should work in different testbenches.
E.g. the HGCROC emulator UVC works as an active (i.e. producing stimuli) component in the ASIC-level testbenches, but it only works as a passive (i.e. monitoring the data received from the HGCROC RTL) component in system-level testbenches.
Designing components for re-use clearly increases their development time but at the same time, it allows implementing them only once, without having to rewrite the same code multiple times with the associated downsides: technical risk due to human errors, subtle mistakes in the implementation, and development and debug time.

Given the similarities between the two ECONs inputs and output stages, both the corresponding UVCs and reference models were written once and re-used across the ASIC-, and system-level testbenches. 
This approach greatly reduced the development times of the verification environment.

\section{Verification of SEE tolerance}
SEE tolerance was considered an additional requirement when setting the verification goals.
As already mentioned, formal verification techniques were used to prove the correct implementation of TMR. 
These techniques have the advantage of being very fast to find bugs (or to prove their absence).
On the other hand, these techniques cannot be universally used: e.g. they have convergence problems with big designs for which only certain parts can be proven to be bug-free. 
They cannot be run on designs with multiple clocks, for which a proven solution still does not exist, and they cannot prove SET tolerance.
Simulation-based techniques~\cite{vrf:sim2023} were used to complement the formal techniques, including:
\begin{itemize}
\item Running SEU injections in fully triplicated nodes and comparing selected ASIC signals against the output of a reference simulation without any fault injection.
\item Running SET injections in triplicated clock and reset trees in post-layout, Gate Level Netlist (GLN) with annotated delays.
\end{itemize}

\section{Runtime of sign-off runs}
To sign-off each of the ECONs the following runs were executed on different DUTs: RTL, triplicated RTL, GLN (with annotated delays, in the three corners) and the fault injection (SEEs) runs.

The verification environment of both ECONs was developed in parallel with the RTL.
Both went through development cycles where bugs were found and subsequently fixed. 
The verification framework is run in regression runs: periodically to validate change requests.
The run time, as one would expect increases as more features are added.
Sign-off runs, i.e. those regressions runs that target the achievement of full coverage of the verification goals, represent a good example of runtime and they are reported in the table~\ref{table:signoff}.

\begin{table}[thbp]
\centering
\caption{Typical run times and number of tests for sing-off runs for each DUT level\label{table:signoff}.
Runtime is calculated on 32 parallel runs.
The regressions were run on machines with two AMD
EPYCTM7302 processors (for a total of 32 cores running at 3 GHz).
ECON-D RTL run includes also the block-level testbench.}
\smallskip
\begin{tabular}{l|cc|cc}
\hline
\multirow{2}*{DUT}     & \multicolumn{2}{c}{Number of tests}     &   \multicolumn{2}{c}{Runtime [hours]} \\
  & ECON-D & ECON-T & ECON-D & ECON-T \\
\hline
RTL & 6000 & 2500 & 33 & 15 \\
TMR with SEU & 1500 & 1500 & 35 & 25\\
GL (SDF) & 1200 & 1200 & 40 & 42\\
GL (SDF) with SEE & 1000 & 1000 & 32 & 35 \\
\hline
\end{tabular}
\end{table}

\section{Conclusions}
The verification strategy developed for both ECON-D and ECON-T was presented.
The final regressions achieved complete functional coverage for the defined operational scenarios.
During the functional verification of ECON-D 216 functional, 67 triplication, and 74 implementation bugs were found.
In ECON-T, 44 functional, 9 triplication, and 14 implementation bugs were found.
The authors attribute the ECON-T lower bug rate to a more mature set of specifications, i.e. a prototype of ECON-T was already existing and partially verified, and the project timeline, i.e. ECON-T was submitted after ECON-D, meaning that the bugs in the common infrastructure were already addressed.
The bugs found vary widely from gaps in the specifications, missing voters, and corner cases in the communication with HGCROC.

Production versions of the two ASICs were submitted for fabrication in December of 2023.
Comprehensive chip testing and radiation characterization have revealed no major issues.
The ASICs are also being integrated in the first detector modules operating together with multiple HGCROCs and lpGBT.

\end{document}